# INTERFEROMETRIC OPTICAL SIGNATURE OF ELECTRON MICROBUNCHING IN LASER-DRIVEN PLASMA ACCELERATORS

A. H. Lumpkin[1], M. LaBerge[2], D. W. Rule[3], R. Zgadzaj[2], A. Hannasch[2], O. Zarini[4,5], B. Bowers[2], A. Irman[4], J. P. Couperus Cabadağ [4], A. Debus[4], A. Köhler[4,5], U. Schramm[4,5], M. C. Downer*[2]

[1]Accelerator Division, Fermi National Accelerator Laboratory, Batavia, IL 60510 USA
[2] Physics Department, Univ. of Texas-Austin, Austin, TX 78712 USA
[3]Silver Spring, Maryland 20904 USA
[4]Institute of Radiation Physics, Helmholtz-Zentrum Dresden-Rossendorf, 01328 Dresden, Germany
[5]Technische Universität Dresden, 01062 Dresden, Germany

## ABSTRACT

We report observations of coherent optical transition radiation interferometry (COTRI) patterns generated by microbunched ~200-MeV electrons as they emerge from a laser-plasma accelerator. The divergence of the microbunched portion of electrons, deduced by comparison to an analytical COTRI model, is ~6x smaller than the ~3 mrad ensemble beam divergence, while the radius of the microbunched beam, obtained from COTR images on the same shot, is < 3 microns. The combined results show that the microbunched distribution has estimated transverse normalized emittance ~0.5 mm mrad.

Key words: laser-plasma accelerator, coherent optical transition radiation, emittance

*downer@physics.utexas.edu

Periodic longitudinal density modulation of relativistic electron beams at optical wavelengths (microbunching) gives rise to coherent light emission in such forms as synchrotron radiation, including the free-electron laser (FEL) [1,2], and optical transition radiation (OTR) in its coherent form (COTR). The latter has been observed from FELs [3,4] and laser-driven plasma accelerators (LPAs) [5,6]. In the first case, the FEL mechanism fundamentally depends on growth of microbunching at the resonant wavelength and its harmonics [7]. In the second case, COTR in some configurations can characterize microbunched portions of the electrons [3-6]. Microbunching in an FEL oscillator was observed indirectly via the buildup of FEL output power to saturation [2]. The first direct time-resolved observation of microbunching in an FEL oscillator [8] used an off-phase final rf accelerator stage to streak a beam modulated at 60 μm wavelength, thereby mapping microbunch arrival time onto energy as displayed in an electron spectrometer. With the advent of self-amplified spontaneous emission (SASE) FELs with a single-pass through a long amplifier chain, FEL light and the electron beam became accessible after each undulator, enabling tracking of FEL power and microbunching. The first measurements of microbunching evolution at visible wavelengths [3,4] used COTR interferometry (COTRI) to track microbunched features uniquely through the exponential gain regime, through saturation, and into post saturation [4]. An analytical model of COTRI fringe patterns showed growth of a microbunched transverse core in the exponential gain regime, and its subsequent reduction after saturation [4, 9]. Subsequently, such experiments were extended to vacuum ultraviolet wavelengths, further benchmarking this model [10, 11]. These experiments foreshadowed the x-ray SASE FELs of today [2, 12-16] by benchmarking the GENESIS simulation code [17] used in their prediction and development [18].

In this Letter, we report new investigations of microbunching in laser-driven plasma accelerators (LPAs) [19], including the first evaluations of the beam emittance of this subset of electrons with COTR techniques. This microbunching is not accessible with betatron x-ray spectroscopy [20], pepper-pot measurements [21], scintillator-based methods [22], or other LPA beam diagnostics [23]. LPAs are currently a major initiative in advanced accelerator technology, with compact FELs [24,25] prominent amongst a growing list of potential applications. Detailed understanding of microbunching is critical to developing LPAs and LPA-based FELs. Past experiments have used COTR to deduce the presence of sub-bunches located in adjacent LPA buckets [26] or of an intrabunch slice of sub-per cent energy spread [27]. Here we use the interference of COTR from 2 tandem foils located downstream of the LPA to deduce the presence of visible-wavelength microbunching within the dominant quasi-monoenergetic component of an electron bunch that was ionization-injected [27, 28] into, and accelerated to ~200 MeV within the leading bubble of a strongly nonlinear LPA. Indeed particle-in-cell simulations have predicted microbunching in LPAs [29, 30]. Such a microbunched fraction might be used to seed an FEL. Our results show that visible-wavelength COTR gain relative to incoherent OTR rivals that obtained previously in a saturated SASE FEL. As a result, COTR is intense enough to distribute to multiple cameras with different frequency filters or imaging modalities on each shot, enabling thorough characterization. For example, one group of cameras detects COTR imaged from the surface of the first foil [hereafter "near-field" (NF) images], which when analyzed using a coherent point-spread function enables beam-size measurements at the LPA exit. Other cameras record the COTR in the focal plane of a collecting lens [hereafter "far-field" (FF) images], thereby measuring the angular distribution of radiation. We observed fringes in the latter data that are consistent with Wartski two-foil COTRI [31]. The first fringes shift outside of the single-foil $1/\gamma$ opening angle of the angular distribution pattern, where $\gamma$ is the relativistic Lorentz factor of the electrons. Fringe visibility indicated microbunched electrons diverged less than the beam-ensemble observed in a downstream electron spectrometer. Experimental results are interpreted using a COTRI model [4,9] to extract electron divergence and pointing from FF data. Combined frequency-filtered beam-size and divergence data yield single-shot, transverse emittance estimates of these microbunched electrons generated within the plasma bubble.

Our experiments used pulses from the DRACO laser (central wavelength 800 nm, energy up to 4 J on target, pulse length 27 fs (FWHM), and peak power 150 TW [28, 32] at Helmholtz-Zentrum Dresden-Rossendorf. These were focused to ~20 µm (FWHM) at the center of a 3-mm-long He gas jet (with 3% Nitrogen and ~0.5-mm-long entrance and exit ramps) to drive LPAs in a self-truncated ionization-injection regime [27] in plasma of density $n_e$ ~3 x $10^{18}$ cm$^{-3}$ [28]. A 75-µm-thick Al laser foil 700 µm from the exit of the jet, tilted ~3º off normal (Fig. 1a), blocked the drive laser pulse. An aluminized Kapton foil located 1 mm downstream blocked **j** x **B** electrons of low energy and associated COTR from the back of the blocking foil [33] as shown in Fig. 1b. Indeed, direct laser excitation of the blocking foil with the gas jet turned off yielded no detectable OTR. With the gas jet turned on, LPA electrons generated forward OTR/COTR from the aluminized back surface of the Kapton foil. Several dozen of these foil pairs were mounted on a 15-cm diameter wheel, which rotated a fresh pair into position for each shot. A 200-μm-thick polished Si wafer oriented at 45° degrees to the beam direction at distance L=18.5 mm downstream of the Kapton foil redirected the foil's OTR/COTR to a 4-cm focal length microscope objective with collection angle 0.14 rad, which relayed it to one group of charge-coupled device (CCD) cameras (12-bit, 3.75-μm square pixels) via beam splitters (Fig. 1a) to record NF images (Fig. 1c). Additionally, the electron bunch generated backward (reflected) COTR from the front surface of the Si wafer (Fig.1a). COTR from these two interfaces, which the microscope objective and an additional 15-cm focal length lens (see Fig. 1a) relayed to another CCD, formed interference fringes in FF images (Fig. 1d). For results reported here, one CCD recorded un-polarized FF images at observation wavelength $\lambda = 633 \pm 5$ nm, while two others recorded NF images through orthogonal linear polarization analyzers [parallel ($x$) and perpendicular ($y$) to the drive laser polarization] at $\lambda = 600 \pm 5$ nm, which were indistinguishable from 633 nm NF images. When the laser focus in the gas jet was adjusted to positions that yielded poly-energetic electron distributions similar to the background in the present LPA output, but with no quasi-monoenergetic peak, we observed OTR signals ~100x weaker than those reported here. Thus, the quasi-monoenergetic peak is the source of reported COTR signals. These signals were intense enough to necessitate neutral density filters to prevent camera saturation. Nevertheless, when we operated the accelerator in different regimes, e.g. by removing the nitrogen dopant and relying on self-injection, we observed strong associated variations in COTR signal strength normalized to accelerated charge. Since foil and laser parameters were unchanged, these variations suggested that the LPA process -- not interaction of electrons with foils or reflected laser fields --- created microbunching responsible for observed COTR. Beam scattering by a foil can *reduce* coherent emission from microbunching when the projected multiple scattering angle exceeds the OTR opening angle $1/\gamma$ [34]. The latter is 2.3 mrad, while the Bethe-Ashkin formula [35] yields a lower value (1.1 mrad) of the former for the Al foil, a higher value (2.6 mrad) for the 45° Si mirror. To corroborate this conclusion, we measured space/angle-integrated forward COTR spectra in a downstream IR-to-UV spectrometer [36]. With the Si wafer temporarily removed, we observed strong IR and visible light down to $\lambda$ ~300 nm. When we re-inserted the Si wafer and placed a new thin OTR foil downstream of it, however, only IR light remained strong. This showed that microbunching responsible for visible COTR survived transit through $\leq 75$ μm Al, but not $\geq 200$ μm/cos45° Si foils. We will present complete COTR spectra and analysis in a planned forthcoming paper.

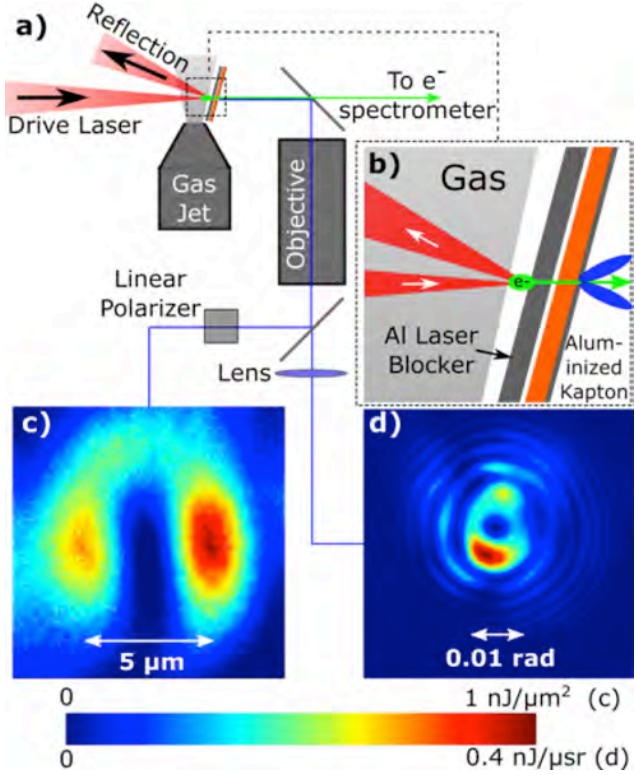

**FIG. 1:** Color online. (a) Schematic setup showing path of LPA drive laser pulse (red), gas jet, OTR foil wheel, Si mirror, and OTR imaging-detection configuration. (b) Detail of gas jet exit, showing electron bunch, drive laser path, blocking foil and aluminized Kapton OTR foil. (c) NF data at 600 nm, imaged from Kapton foil to CCD through linear polarization analyzer, showing two point-spread function lobes. (d) FF data at 633 nm for same shot, recorded with compound lens (consisting of microscope objective + auxiliary single-element lens) and no polarization analyzer, showing interference fringes. Angular resolution is 0.35 ± 0.05 mrad/pixel, or ~6 pixels/fringe at the eighth fringe. Color legend applies to both (c) and (d).

Currents induced when a charged particle beam enters and exits a foil generate, respectively, forward and backward optical transition radiation [37-39]. The backward radiation cone of half angle $1/\gamma$ is generated around the specular reflection direction while forward radiation is generated in a $1/\gamma$ cone around the beam direction. Thus, the configuration in Fig. 1 generates OTR at 90° to the beam direction, enabling minimally invasive OTR detection and imaging. Interference of forward OTR from the aluminized Kapton with the backward OTR from the silicon reflector produces fringes with peak maxima at p = 1/2, 3/2, 5/2, etc., related to foil separation L by $L = p\gamma^2\lambda$. Choosing L = 18.5 mm enabled focusing near-field optics on the first foil, while still providing good fringe contrast.

The number $W_1$ of OTR photons that a single electron generates per unit frequency $\omega$ per unit solid angle $\Omega$ is

$$\frac{d^2W_1}{d\omega d\Omega} = \frac{e^2}{\hbar c} \frac{1}{\pi^2 \omega} \frac{(\theta_x^2 + \theta_y^2)}{(\gamma^{-2} + \theta_x^2 + \theta_y^2)^2} \tag{1}$$

where $\hbar$ is Planck's constant/$2\pi$, $e$ is the electron charge, $c$ is the speed of light, and $\theta_x$ and $\theta_y$ are radiation angles [9]. The black curve in Fig. 2a shows the single-foil OTR angular distribution with peaks at $\theta \equiv \theta_x = \theta_y = 1/\gamma \sim 2.3$ mrad, generated by a 200 MeV, N-electron bunch with divergences in both planes of 0.2 mrad. For $N$ particles, $N_B$ of which are microbunched, the coherence function $J(k)$ becomes involved, and for two foils an interference function $I(k)$ comes into play. The spectral angular distribution function then becomes

$$\frac{d^2W}{d\omega d\Omega} = |r_{\parallel,\perp}|^2 \frac{d^2W_1}{d\omega d\Omega} I(k)J(k) \tag{2}$$

where $r_{\parallel,\perp}$, is the reflection coefficient of the second foil for parallel and perpendicular polarization components, respectively. $I(k)$ is given by [31]

$$I(k) = 4\sin^2[\frac{kL}{4}(\gamma^{-2} + \theta_x^2 + \theta_y^2)] \tag{3}$$

where $k = |\mathbf{k}| = 2\pi/\lambda$, and we used a small-angle approximation. The dashed red curve in Fig. 2a shows the corresponding two-foil OTR angular distribution for $L$= 18.5 mm and $\lambda$= 633 ± 5 nm. At $\gamma$ values of interest, intensity asymmetry of the first peaks of the parallel polarization component in Fig. 2a are negligible (see Eq. 1 ref. [38]). Strong fringe modulation is seen in this example where a Gaussian beam divergence of 0.2 mrad was convolved with (2). The coherence function can be defined as

$$J(\mathbf{k}) = N + N_B(N_B - 1)|H(\mathbf{k})|^2 \qquad (4)$$

where

$$H(\mathbf{k}) = \frac{\rho(\mathbf{k})}{Q} = g_x(k_x)\, g_y(k_y)\, F_z(k_z) \qquad (5)$$

is a product (where we assume separability of $\rho(\mathbf{k})$) of Fourier transforms $g(k_i)=exp(-\sigma_i^2 k_i^2/2)$ of the transverse charge form factors ($i=x,y$), with $k_i \approx k\theta_i$, and $F_z(k_z) = exp(-\sigma_z^2 k_z^2/2)$ of the longitudinal form factor, with $k_z \sim k$ for $\theta <<$ 1. Q is the total charge of one electron bunch. $J(\mathbf{k})$ reduces to $N$ when $N_B = 0$.

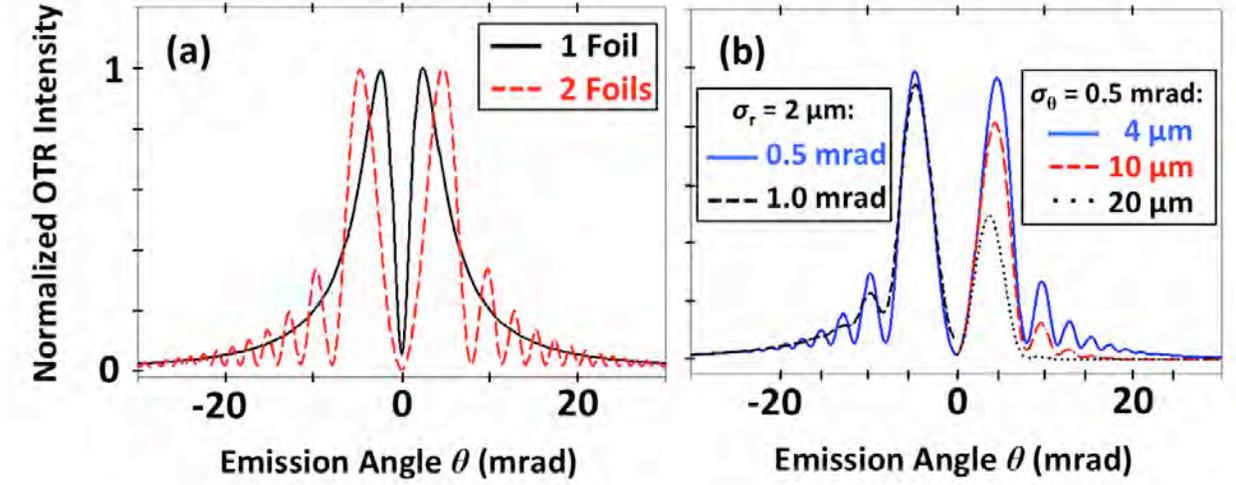

**FIG. 2**. Color online. (a) Analytical model calculations of OTR angular distributions at λ=633 ± 5 nm generated by 200 MeV electron bunch with divergences $\sigma_{x'y'}$ = 0.2 mrad: in a single-foil (solid black curve) source and in a Wartski two-foil interferometer with L= 18.5 mm (dashed red curve). (b) Analytical model calculations showing dependence of COTRI fringe visibility on beam divergence and radius. For negative angles, the beam-divergence effect is shown for 0.5 mrad (solid blue line) and 1.0 mrad (dashed black line, reduced modulation) for 2-μm beam radius. For positive angles, Gaussian beam size effects for 4, 10 and 20 μm (all for 0.5-mrad divergence) are shown. For the larger beams, the outer fringes are suppressed.

Figure 2b illustrates how transverse charge form factors determine COTR gain vs. $\theta$, for an azimuthally symmetric ($\sigma_r = \sigma_x = \sigma_y$) beam of micro-bunching fraction $f_B = N_B/N = 0.01$. The left-hand side ($\theta < 0$) shows effects of microbunched beam divergences ($\sigma_\theta$ = 0.5, 1.0 mrad) for fixed $\sigma_r$ = 2 μm, the right-hand side ($\theta > 0$) effects of microbunched beam radii ($\sigma_r$ = 4, 10, and 20 μm) for fixed $\sigma_\theta$ = 0.5 mrad. 8 fringes are enhanced for the 2-μm, 0.5-mrad case, but only one for the 20-μm, 0.5-mrad case. Past COTRI studies of SASE FELs involved beam radii 25 μm < $\sigma_{x,y}$ < 200 μm, and yielded fringes out to $\theta_{x,y} \sim$ 5 mrad [4,9] depending on microbunched core size. LPA beams are often an order of magnitude smaller, so coherent enhancements can be proportionately larger and can occur out to $\theta_{x,y}$ > 30 mrad. Generally, $\sigma_\theta$, $\sigma_r$, electron energy bandwidth $\Delta\gamma$, and optical detection bandwidth $\Delta\lambda$ can all affect fringe visibility in a FF angular distribution pattern. Our chosen bandwidth filter ($\Delta\lambda$ = 10 nm), however, minimizes the bandwidth effect, while COTRI intensity and fringe visibility are nearly $\gamma$−independent for $\gamma >$ 400 (see Supplementary Material for details). Thus we can relate observed fringe number directly to $\sigma_\theta$ and $\sigma_r$ of the contributing portion of the beam and observed COTR intensity directly to $f_B$. Consequently, observation of ≥ 3 fringes implies sub-mrad divergence and few-μm beam radius with this bandpass filter. Simultaneously, by calibrating integrated signal on the FF camera, we determine $J(k)$, and thus COTR gain ($\sim N_B^2/N$ for $N_B \gg 1$) over incoherent OTR and $f_B$. We find $N = 1.47 \times 10^9$ electrons in the quasi-monoenergetic peak (i.e. $eN = 235$ pC) and $N_B = 1.8 \times 10^7$ (i.e. $eN_B \approx 3$ pC), implying $f_B \approx 0.013$ and $N_B^2/N \approx \sim240,000$. These estimates took into account the single-electron, single-foil OTR source energy [40] and the measured charge in the quasi-monoenergetic peak based on a calibrated LANEX screen [41] in the spectrometer on the same shot (see Supplementary Material for details). Thus coherent enhancements dominate over incoherent OTR for these conditions.

The FF image in Fig. 1d has 8 to 9 visible fringes. The azimuthal asymmetry in fringe intensity probably results from transverse distortions of the COTR source from a Gaussian charge distribution. Here, to simplify analysis, we averaged the data in Fig. 1d azimuthally. We then assess the resulting fringe pattern (Fig. 3a, black curve) by comparison to analytical results in Fig. 2b. Since the outermost fringes (Fig. 3a, inset) are more sensitive to $\sigma_\theta$ than the first two fringes, whereas amplitudes of the latter are more sensitive to neglected non-Gaussian features of the beam shape, we analyzed $\sigma_\theta$ quantitatively by fitting azimuthally symmetric COTRI model calculations for various $\sigma_\theta$ to the 4$^{th}$ thru 9$^{th}$ azimuthally averaged fringes. The dashed red curve in Fig. 3a shows the complete best-fit COTRI curve. Full analysis yields $\sigma_\theta = 0.48^{+.10}_{-.09}$ mrad (see Supplementary Material for details). This is significantly smaller than the ensemble divergence of 3 mrad measured at the electron spectrometer without foils present [28]. Calculated and measured fringe peak positions agree well, and the number of fringes detectable (8 to 9) in the azimuthal average can only be explained with a sub-mrad divergence value.

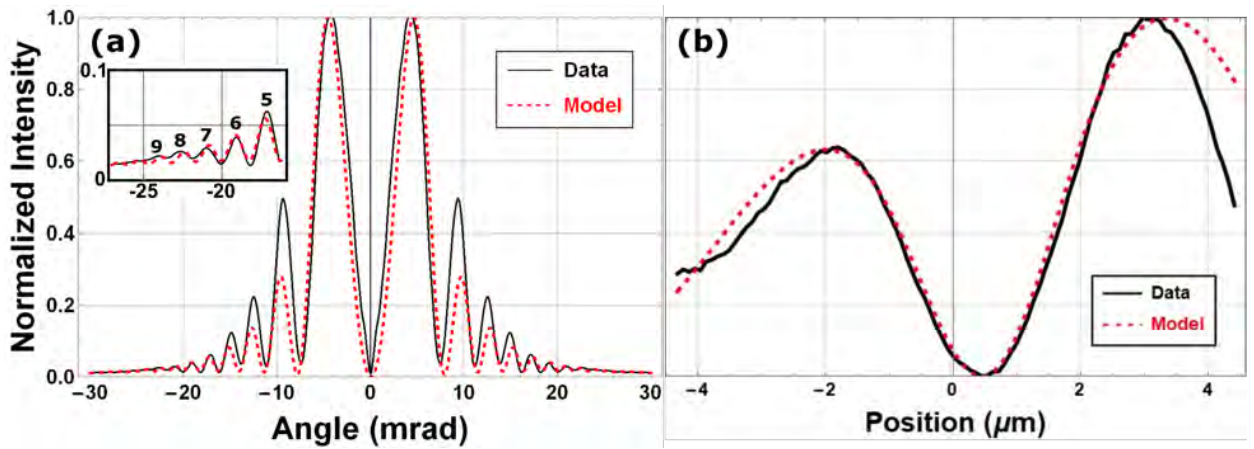

**FIG. 3**: Color online. **(a)** Azimuthally averaged COTRI fringes vs. $\theta$ from Fig. 1d (solid black curve), with outermost fringes magnified in inset, compared to analytical COTRI model using $\sigma_\theta = 0.5$-mrad (dotted red curve). **(b)** x-projection of NF image in Fig. 1c from same shot (solid black curve), with COTR model using $\sigma_x = 2.75$-μm (dotted red curve).

This number of fringes requires $\sigma_{x,y} < 6$ μm based on modeling results in Fig. 2b. Analysis of linear polarized NF images (e.g. Fig. 1c) confirms this conclusion. Fig. 3b compares a COTR model calculation for $\sigma_x = 2.75$ μm (red dotted) to a y-averaged version of x-polarized NF data in Fig. 1c (Fig. 3b, black curve). Here, calculated curves take into account the finite optical collection angle via Eq. 26 of Ref. [42]. The left-right asymmetry of the NF data, again, probably results from distortions of the beam charge profile from a Gaussian shape, which we introduced into the model with an empirical skew parameter. Full analysis yields $\sigma_x = 2.75^{+0.45}_{-0.30}$ μm (see Supplementary Material for details). When combined with the $0.48^{+0.10}_{-0.09}$ mrad x-divergence, we obtain an estimated normalized emittance of $\varepsilon_{nx} = 0.53^{+0.14}_{-0.11}$ mm mrad (rms) for this subset of the beam. Similar values were obtained for y-polarized NF data. These values were obtained at distance $z \approx 2.2$ mm from the end of the constant-$n_e$ plateau of the gas jet. Interpreting the latter as the accelerator exit, and extrapolating $\sigma_x = 2.75^{+0.45}_{-0.30}$ back to $z = 0$ using $\sigma_\theta = 0.48^{+.10}_{-.09}$ mrad yields estimated rms beam radius $\sigma_x(z=0) \approx 1.5$ μm at the accelerator exit for this shot. This value is consistent with the range of rms beam radii inside, and near the end of, this accelerator determined independently by analyzing betatron x-ray spectra [43], results which will be published separately.

To illustrate wider COTRI diagnostic possibilities for LPA beams with more complex structure, Fig. 4a shows FF data without linear polarizer for a different shot. Here, a dark node runs vertically through the interference pattern, a feature not seen in Fig. 1d, nor in most shots.

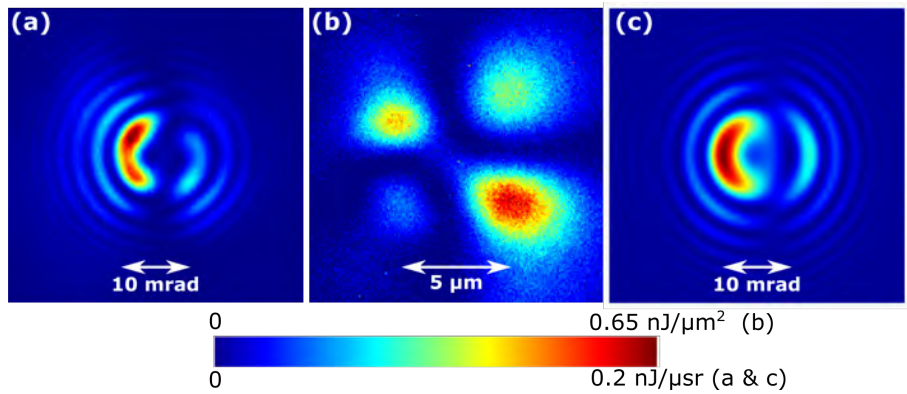

FIG. 4: Color online. (a) COTR angular distribution pattern for shot with vertical dark band due to interference of two beamlets. The intensity display scale maximum was clamped to show the outer fringes more clearly. (b) The y-polarized NF image on the same shot, showing two double lobe patterns separated by ~6 μm. (c) a COTRI image reconstructed from a two-beamlet model (2) with delta phase of 0.75 π and delta trajectory angle of 2 mrad. Color legend applies to all panels.

The corresponding NF image (Fig. 4b) shows two pairs of point-spread-function lobes separated by ~6 μm along x, instead of one pair as in Fig. 1c, suggesting that two beamlets emerged side-by-side from the LPA. Such bi-modal beam distributions along the laser polarization have been observed in PIC simulations of bubble-regime LPAs (see Fig. 6 of Ref. [44]). Fig. 4c shows a reconstruction of the main qualitative features of Fig. 4a by modeling two beamlets having a phase difference of 0.75 π with angular trajectories differing by 2 mrad, each with $\sigma_\theta$ = 0.6 mrad. This example thus illustrates that FF COTRI patterns contain signatures of the phase and trajectory of multiple microbunched beamlets, when present. Combined FF and NF data indicate single-beamlet normalized emittance similar to that obtained from the data in Fig. 3.

In summary, we have identified visible wavelength microbunching in electron beamlets accelerated in LPAs with concomitant COTR gain over incoherent OTR exceeding $10^5$. We have identified COTR interference fringes in FF data and reproduced their main features with an analytical COTRI model originally developed for SASE-FEL-induced microbunching, but here applied to LPAs for the first time. Implied divergences and single-shot emittances of microbunched beamlets are significantly less than those of the ensemble of electrons. The percent-scale microbunching observed here, if transportable [6], could seed a SASE FEL startup more readily than weaker longitudinal-space-charge-instability microbunching [45] and/or provide beam diagnostic information from undulator radiation. Future directions include exploring the wide variations in COTR intensity that we have observed in different LPA operating regimes, understanding their connection to microbunching mechanisms, and developing strategies for diagnosing, tuning and optimizing microbunching over a wide range of wavelengths.

A.H.L. acknowledges discussions with K.-J. Kim (ANL) on microbunching in FELs and R. Thurman-Keup (FNAL) on COTRI. This manuscript has been authored by Fermi Research Alliance, LLC under Contract No. DE-AC02-07CH11359 with the U.S. Department of Energy, Office of Science, Office of High Energy Physics. University of Texas authors acknowledge support from DoE grant DE-SC0011617, and M.C.D. from the Alexander von Humboldt Foundation with sponsorship from R. Sauerbrey. Helmholtz-Zentrum Dresden-Rossendorf authors acknowledge support by the Helmholtz Association under program Matter and Technology, topic Accelerator Research and Development.

# References

[1] J. M. J. Madey, *J. Appl. Phys*. **42,** 1906 (1971).

[2] Kwang-Je Kim, Zhirong Huang, and Ryan Lindberg, *Synchrotron Radiation and Free-Electron Lasers: Principles of Coherent X-ray Generation* (Cambridge University Press, 2017).

[3] A.H. Lumpkin, R. Dejus, W.J. Berg, M. Borland, Y.C. Chae, E. Moog, N.S. Sereno, and B.X. Yang, *Phys. Rev. Lett.* **86,** 79 (2001).

[4] A. H. Lumpkin *et al.*, *Phys. Rev. Lett*. **88**, 234801 (2002).

[5] Y. Glinec, J. Faure, A. Norlin, A. Pukhov, and V. Malka, *Phys. Rev. Lett*. **98**, 194801 (2007).

[6] C. Lin *et al.*, *Phys. Rev. Lett*. **108**, 094801 (2012).

[7] Kwang-Je Kim, *Nucl. Instr. and Meth. Phys. Res. A* **250**, 396 (1986).

[8] Kenneth N. Ricci and Todd I. Smith, *Phys. Rev. ST-Accel. and Beams* **3**, 032801 (2000).

[9] D.W. Rule and A.H. Lumpkin, "Analysis of coherent optical transition radiation interference patterns produced by SASE-induced microbunches," in *Proceedings of the 2001 Particle Accelerator Conference*, vol. 2. (IEEE, Piscataway, NJ, 2001), pp. 1288-1290.

[10] A.H. Lumpkin, R.J. Dejus and D.W. Rule, "First direct comparisons of a COTRI analytical model to data from a SASE FEL at 540-nm, 265-nm, and 157-nm," in *Proceedings, 26$^{th}$ International Conference, Free Electron Laser 2004* (Comitato Conferenze Elettra, Treiste, Italy, 2004), pp. 519-522.

[11] A.H. Lumpkin, Y.-C. Chae, R.J. Dejus, M. Erdmann, J.W. Lewellen, and Y. Li, "Use of VUV Imaging to Evaluate COTR and Beam-Steering Effects in a SASE FEL at 130 nm," in *Proceedings, 26$^{th}$ International Conference, Free Electron Laser 2004* (Comitato Conferenze Elettra, Treiste, Italy, 2004), pp. 523-526.

[12] J. Arthur *et al*., "Linac Coherent Light Source (LCLS) Conceptual Design Report", SLAC Report SLAC-R-593 (2002).

[13] JADRI/Spring-8 and Harima RIKEN, "SPring-8 Compact SASE Source Conceptual Design Report," eds. T. Tanake and T. Shintake (2005).

[14] H.-S. Kang, K.-W. Kim, and I.-S. Ko, "Current Status of the PAL-XFEL Project" in *Procceedings, International Particle Accelerator Conference 2014*, (JACoW Publishing, Dresden, Germany, 2014), pp. 2897-2899.

[15] R. Brinkman, B. Faatz, K. Flöttmann, J. Rossbach, J.R. Schneider, H. Schult-Schrepping, D. Trines, T. Tschentscher, and H. Weise, "TESLA XFEL: First stage of the X-ray Laser Laboratory", DESY, Report TESLA FEL2002-09 (2002).

[16] Paul Scherrer Institute (PSI), "Swiss FEL Conceptual Design Report", ed. R. Ganter (2010).

[17] S. Reiche, *Nucl. Instr. Meth. Phys. Res. A* **429**, 243 (1999).

[18] S.V. Milton *et al.*, *Science* **292**, 2037 (2002).

[19] T. Tajima and J.M. Dawson, *Phys. Rev. Lett.* **43**, 267 (1979).

[20] G.R. Plateau *et al.*, *Phys. Rev. Lett*. **109**, 064802 (2012).

[21] E. Brunetti *et al.*, *Phys. Rev. Lett.* **105**, 215007 (2010); A. Cianchi *et al., Nucl. Instr. Meth. Phys. Res. A* **720**, 153 (2013).

[22] S. K. Barber *et al.*, *Phys. Rev. Lett.* **119**, 104801 (2017).

[23] M. C. Downer, R. Zgadzaj, A. Debus, U. Schramm, and M.C. Kaluza, *Rev. Mod. Phys*. **90**, 035002 (2018).

[24] F. Gruener *et al.*, *Appl. Phys. B* **86**, 431 (2007); T. André *et al*., *Nat. Commun*. **9**, 1334 (2018).

[25] K. Nakajima, *Nat. Phys*. **4**, 92 (2008).

[26] O. Lundh, C. Rechatin, J. Lim, V. Malka, and J. Faure, *Phys. Rev. Lett.* **110**, 065005 (2013).

[27] M. Mirzaie *et al.*, *Sci. Rep.* **5**, 14659 (2015).

[28] J. Couperus *et al.*, *Nat. Commun.* 8, 487 (2017); A. Irman, J. P. Couperus, A. Debus, A. Köhler, J. M. Krämer, R. Pausch, O. Zarini, and U. Schramm, *Plasma Phys. Control. Fusion* **60**, 044015 (2018).

[29] X. L. Xu *et al.*, *Phys. Rev. Lett.* **117**, 034801 (2016).

[30] K. Németh, B. Shen, Y. Li, H. Shang, R. Crowell, K.C. Harkay, and J.R. Cary, *Phys. Rev. Lett.* **100**, 095002 (2008).

[31] L. Wartski, S. Roland, J. Lasalle, M. Bolore, and G. Fillippi, *J. Appl. Phys.* **46**, 3644 (1975).

[32] U. Schramm *et al.*, in *Proceedings, International Particle Accelerator Cconference 2017* (IOP Publishing, Copenhagen, Denmark, 2017), IOP Conf. Series: *Journal of Physics: Conf. Series* **874**, 012028.

[33] H. Popescu, D.D. Baton, and F. Amiranoff, *Phys. Plasmas* **12**, 063106 (2005).

[34] A. Murokh, E. Hemsing, and J. Rosenzweig, "Multiple Scattering-Induced Mitigation of COTR Emission from Microbunched Electron Beams" in *Proceedings, Particle Accelerator Conference 2009*, (JACoW Publishing, Vancouver, BC, Canada, 2009), TH6REP021.

[35] E. Segrè *et al.*, *Experimental Nuclear Physics, Vol. I* (John Wiley and Sons, NY, 1953), p.285.

[36] O. Zarini *et al.*, "Advanced Methods for Temporal Reconstruction of Modulated Electron Bunches," in *Proceedings, Advanced Accelerator Concepts Workshop 2018*, (IEEE, Breckenridge, CO, USA, 2018).

[37] V. L. Ginzburg and I. M. Frank, *Sov. Phys. JETP* **16**, 15 (1946).

[38] D. W. Rule, *Nucl. Instrum. Meth. Phys. Res. B* **24/25**, 901-904 (1987).

[39] A. H. Lumpkin, in *Proceedings, Accelerator Instrumentation Workshop 1990*, (Batavia, IL, 1991) *AIP Conf. Proceedings* **229**, pp. 151-179.

[40] C.B. Schroeder, E. Esarey, J. van Tilborg, and W.P. Leemans, *Phys. Rev. E* **69,** 016501 (2004).

[41] T. Kurz *et al.*, *Rev. Sci. Instrum.* **89**, 093303 (2018).

[42] M. Castellano and V. A. Verzilov, *Phys. Rev. ST-Accel. Beams* **1**, 062801 (1998).

[43] A. Köhler, R. Pausch, J. P. Couperus Cabadağ, O. Zarini, J. M. Krämer, M. Bussmann, A. Debus, U. Schramm and A. Irman, "Minimizing betatron coupling of energy spread and divergence in laser-wakefield accelerated electron beams," arXiv 1905.02240v1 (2019).

[44] A. H. Lumpkin, R. Crowell, Y. Li, and K. Nemeth, "A Compact Electron Spectrometer for an LWFA," in *Proceedings, 29th International Conference, Free Electron Laser 2007*, (Curan Associates, Inc.; Novosibirsk, Russia, 2007) WEAAU05.

[45] A.H. Lumpkin, R.J. Dejus, and N.S. Sereno, *Phys. Rev. ST-Accel. and Beams* **12**, 040704, (2009).

**Supplementary Material** (Lumpkin *et al.*, "Interferometric optical signature...")

**1. *Effect of detection bandwidth on COTRI fringe visibility.*** Generally in a FF angular distribution pattern, the optical bandwidth of the filter through which the radiation is detected influences fringe visibility. Fig. S1 illustrates this effect for the case of COTR generated in two tandem films separated by $L = 18.5$ mm by a Gaussian electron bunch of energy $E_e = 200$ MeV with $\sigma_r = 2\mu$m, $\sigma_\theta = 0.5$ mrad. Calculated COTRI patterns are shown for $\lambda = 633$ nm and detection bandwidths $\Delta\lambda_{FWHM}$ = 0, 10 and 40 nm. The patterns are calculated by convolving the field from a single wavelength [Eq. 8 of Wartski, J. Appl. Phys. (1975)] with a Gaussian wavelength distribution of the given bandwidth.

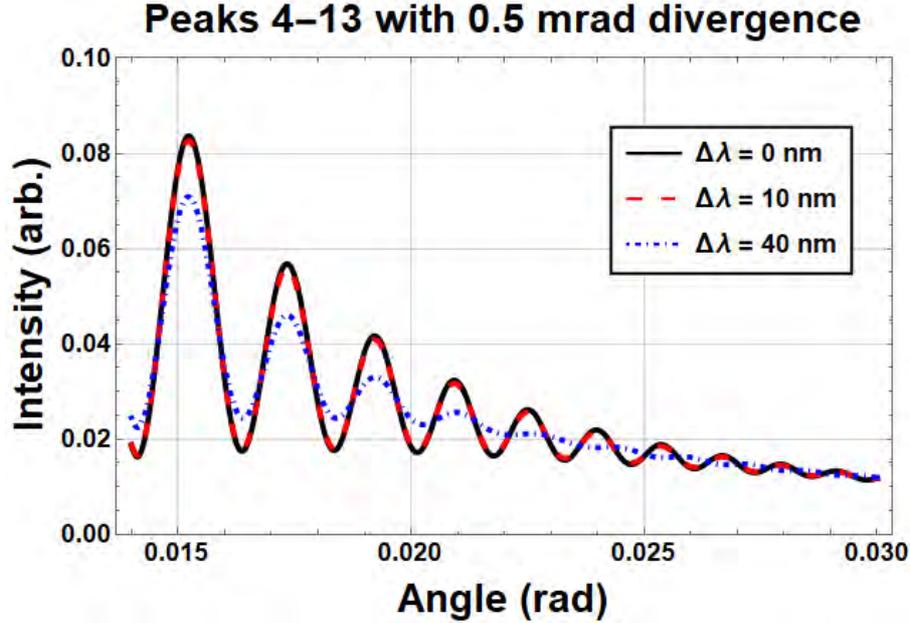

**Fig. S1**. Calculated COTRI angular distributions generated by electron bunch of energy $E_e = 200$ MeV, radius $\sigma_r = 2\mu$m, divergence $\sigma_\theta = 0.5$ mrad in films separated by $L = 18.5$ mm, detected through 633 nm optical bandpass filters of $\Delta\lambda_{FWHM}$ = 0 (solid black), 10 (dashed red) and 40 (dotted blue).

The results show that for $\Delta\lambda_{FWHM} = 10$ nm, the bandwidth chosen for experimental results presented in the main text, fringe visibility is close to that of ideal monochromatic detection. Thus observed fringe visibility can be interpreted purely in terms of e-beam parameters. For $\Delta\lambda_{FWHM} = 40$ nm, on the other hand, significant fringe visibility is lost because of detection bandwidth alone.

**2. *Calibration of COTR gain, microbunching fraction.*** COTR gain $N_b^2/N$ and microbunching fraction $N_b/N$ were determined by calibrating the integrated optical energy deposited in the FF detector, the throughput of the optical path from OTR foils to detector, and the total charge Q within the quasi-monoenergetic 200 MeV peak of the LPA output that was detected in the electron spectrometer. The integrated FF signal in Fig. 1d corresponds to deposited optical energy 7.4 pJ at $\lambda = 633$ nm. This is based on an in-house calibration of the CCD camera's detection efficiency using a HeNe laser of independently measured power, and spot size similar to that of the detected signal. Measured and/or published transmission and reflectivity coefficients of individual elements --- microscope objective, beam splitters, silicon wafer, bandpass filter and neutral density filters --- of the aluminized-Kapton-foil-to-detector relay line yielded overall throughput 0.013% within the bandwidth $628 < \lambda < 638$ nm, implying 58 nJ generated within this bandwidth at the foil. A single 200 MeV electron produces $8.1 \times 10^{-14}$ nJ of radiation within this bandwidth and the solid angle of our collection system at one foil [Eq. 1 of Schroeder, Phys. Rev. 2004] or twice this at two foils, smaller by a factor $3.6 \times 10^{14}$ than the generated energy. For predominantly coherent OTR, we conclude $N_b = [3.6 \times 10^{14}]^{1/2} \approx 1.8 \times 10^7$ electrons, or charge $eN_b \approx 3$ pC. The total charge $Q = Ne = 235$ pC, corresponding to $N = 1.47 \times 10^9$ electrons, in the quasi-monoenergetic peak was determined from the intensity of light emission that it stimulated in a calibrated luminescent screen (Lanex) located at the detection plane of an electron spectrometer. Absolute calibration of the Lanex screen is described in detail in Kurz et al., *Rev. Sci. Instrum.* **89**, 093303 (2018). Thus $N_b/N = 0.013$ and $N_b^2/N = 2.4 \times 10^5$.

**3.** *Analysis of COTRI data.* To compare the COTRI data in Fig. 1d with the model, we averaged over azimuthal variations in the data. To implement this average, we first fit a circle to the outermost visible fringe in Fig. 1d. From this we determined the center of the interference pattern. The data in Fig. 1d was then averaged over radial integration paths to produce the azimuthally averaged data shown by the black curve in Fig. 3a, where it is plotted against the emission angle $\theta$. The common center of the circle was fine-tuned by varying its location within a 7x7-pixel area around the central minimum of the interference pattern and identifying the pixel that maximized visibility of the outermost fringes. The resulting plot provided the basis for further analysis, which proceeded in the following logical sequence:

*a. Insensitivity of COTRI fringes to $\gamma$.* Eq. (3) shows that there is a $\gamma$-dependent phase shift in the COTRI fringe intensity. In our experiments, COTR originates mainly from electrons in the quasi-monoenergetic peak, for which $400 < \gamma < 500$. Fig. S2a plots COTRI intensity generated in two tandem films separated by $L = 18.5$ mm by Gaussian electron bunches of $\gamma = 300, 400$ and $500$ with $\sigma_r = 2\mu m$, $\sigma_\theta = 0$ mrad. A small phase shift is visible for $\gamma = 300$ (dashed blue curve), but intensity and fringe visibility become nearly $\gamma$-independent for $\gamma = 400, 500$ and higher. Consequently fringe visibility in the range of our experiments is unaffected by small electron energy broadening. Fig. S2b shows this explicitly by plotting COTRI intensity at $\gamma = 400$ for bandwidths $\Delta\gamma = 0, 50$ and $250$. The first two plots are indistinguishable from each other. Thus the ~10% energy bandwidth of the quasi-monoenergetic electrons in our experiment does not affect fringe visibility, which therefore is determined by $\sigma_\theta$ alone.

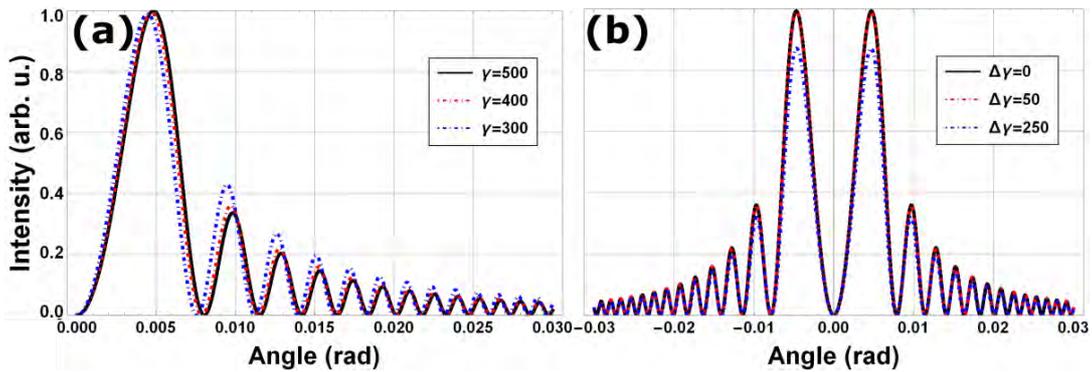

**Fig. S2**. Dependence of COTRI fringes on $\gamma$. (a) Calculated COTRI intensity for three different values of $\gamma$; (b) calculated COTRI intensity at $\gamma = 400$, for three bandwidths of the electron energy distribution.

*b. Determination of divergence $\sigma_\theta$ and its uncertainty from fringe visibility.* Since COTRI fringe visibility is insensitive to $\sigma_r$ in the range $\sigma_r < 4$ μm (see main text, Fig. 2b, positive angles), we fixed the beam radius at $\sigma_r \approx 2$ μm for purposes of fitting azimuthally averaged COTRI data in Fig. 3a with $\sigma_\theta$ as the sole variable. As a preliminary step, we subtracted a $\theta$-dependent background from the data (see Fig. S3a) and obtained the result in Fig. S3b.

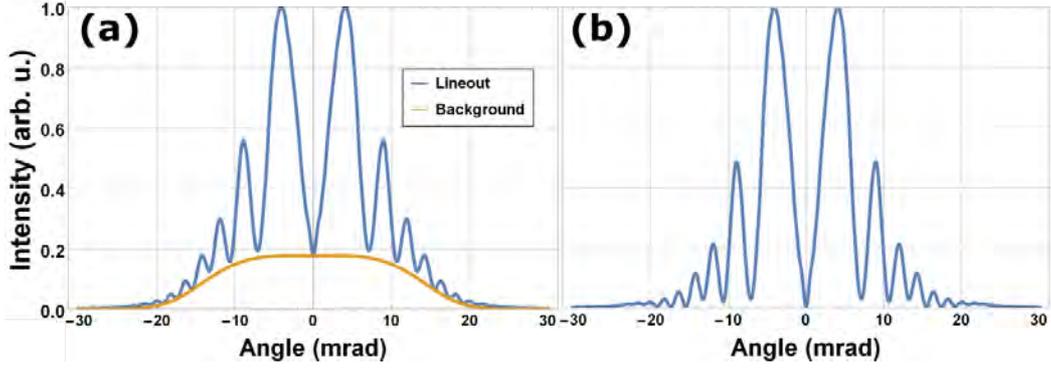

**Fig. S3**. Correction of COTRI data for sub-unity modulation depth. (a) Normalized, azimuthally averaged COTRI data (blue), and background (orange curve) attributable to optical imperfections and finite pixel size. (b) Re-normalized, background-subtracted COTRI data used for quantitative comparison with COTRI model.

The less than unity modulation depth in the raw data results from finite pixel size, optics misalignment, aberrations, and other imperfections in the imaging system. The background-subtracted data was then fit to a family of model COTRI curves with variable $\sigma_\theta$. Fig. S4 shows results of this fit.

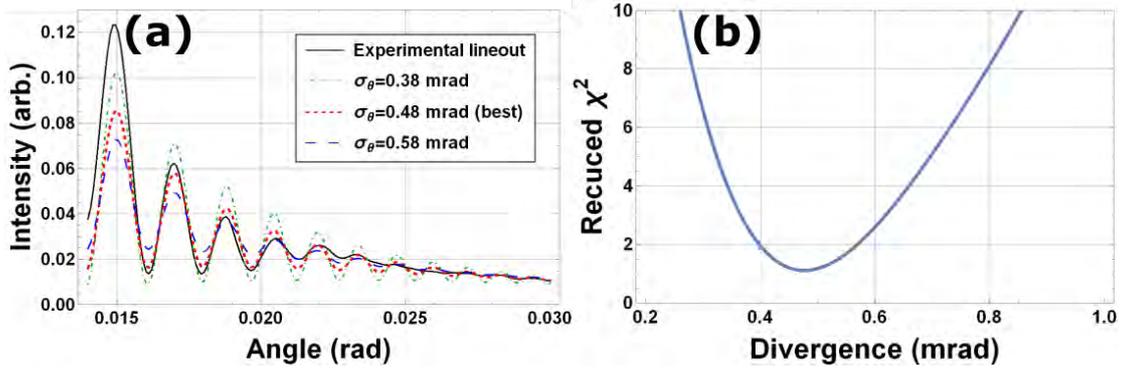

**Fig. S4**. Analysis of wide-angle FF COTRI data to obtain $\sigma_\theta$. (a) Normalized, azimuthally averaged COTRI fringes (black, from Fig. 3a, main text), and modeled fringes for $\sigma_\theta$ = 0.38 (green), 0.48 (red) and 0.58 (green) mrad. Theory curves are offset vertically downward by 0.02 units for easier comparison with data. (b) Reduced $\chi^2$ value of fit (vertical axis) vs. $\sigma_\theta$, yielding $\sigma_\theta = 0.48^{+.10}_{-.09}$ mrad.

The outer fringe visibility is more sensitive to beam divergence than the inner fringes. Also, due to coherence effects, the inner fringes are more sensitive to high spatial frequency features within the transverse momentum distribution than the outer fringes [Eq. 8 of Wartski, J. Appl. Phys. (1975)]. We chose to fit our model to the signal beyond the third peak, thereby selecting for the region that greatest sensitivity to total bunch divergence. The $\chi^2$ value of the fit is minimized when $\sigma_\theta$ = 0.48 mrad (see Fig. S4b). We bounded the uncertainty range at $\sigma_\theta$ values (0.39 and 0.58) that yielded twice the minimum value of $\chi^2$, yielding $\sigma_\theta = 0.48^{+.10}_{-.09}$ mrad. If we fit fringes at a higher or lower cutoff angle, this $\chi^2$ best fit could vary by a few tenths of mrad, but remains below 0.7 mrad regardless of this choice.

**4. *Analysis of near-field COTR data.*** We determined bunch size $\sigma_r$ and its uncertainty by comparing polarized NF data (e.g. Fig. 1c of main text) to predictions of a COTR model, using $\sigma_r$ as a variable fitting parameter. We first integrated polarized NF signals along a direction (*y*) perpendicular to the polarization analyzer's transmission axis, to obtain a double-lobed intensity pattern along *x* (Fig. 3b, main text). We model this pattern by superposing TR *fields* from individual electrons in the bunch. Electron energy and the numerical aperture of the imaging optic determine each electron's imaged TR field pattern, according to Eq. 26 of Castellano Phys. Rev. ST Accel. Beams 1998. Because the observed intensity pattern had lobes of unequal height (see Fig. 3b), we modeled the beam as a skewed normal distribution

$$f(x) = \frac{1}{\omega\sqrt{2}} e^{-\frac{x^2}{2\omega^2}} \left(1 + erf\left(\frac{\alpha\, x}{\omega}\right)\right)$$

where $\omega$ is the width parameter, $\alpha$ the shape parameter and $erf(x)$ is the Gauss error function. The standard deviation of this distribution is $\sigma_x = \omega\sqrt{1 - \frac{2}{\pi}\frac{\alpha^2}{1+\alpha^2}}$, which for $|\alpha| < 0.4$ becomes approximately equal to $\omega$: i.e. $\sigma_x \approx \omega$. The ratio of lobe heights proved sensitive to the skew parameter $\alpha$, but insensitive to the distribution width $\sigma_x$. Thus we first fit the observed lobe height ratio (0.64 for the data in Fig. 3b), yielding $\alpha = 0.15$. As Fig. S5ba shows, this determination was nearly independent of $\sigma_x$. Using this value for $\alpha$, we then generated COTR patterns for varying $\sigma_x$. Fig. S5c shows examples for $\sigma_x$ = 2.45, 2.75 and 3.2 µm. We found that $\chi^2$ was minimized at $\sigma_x$ =2.75 µm, and that bounding $\sigma_x$ values 2.45 and 3.2 µm yielded twice the minimum $\chi^2$ (see Fig. S5b). Thus we cite rms beam x-radius $\sigma_x = 2.75^{+0.45}_{-0.30}$ at position $z \approx 2.5$ mm from the exit of the accelerator, where the TR foil was located.

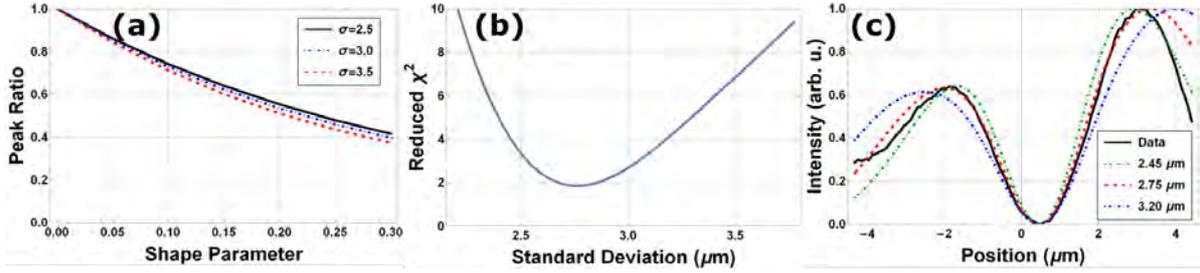

**Fig. S5**. Analysis of NF COTR data to obtain $\sigma_x$. **(a)** Dependence of lobe height ratio (vertical axis) on skewness parameter $\alpha$ (horizontal axis) and $\sigma_x$ (legend). **(b)** $\chi^2$ value of fit (vertical axis) vs. $\sigma_x$ (horizontal axis), yielding $\sigma_x$ =2.75$^{+0.45}_{-0.30}$. **(c)** Normalized, y-averaged NF COTR data (solid black curve, Fig. 3b, main text), and modeled COTR pattern for $\sigma_x$ = 2.45 µm (green dotted curve), 2.75 µm (red dashed), and 3.2 µm (blue dotted).